# Surface Metrology of Cerebral Arteries Luminal Surface

Jennifer Buoncore[1], Alexis Sobecki[1], Brian Beatty, PhD[1*]

[1]NYIT College of Osteopathic Medicine, Northern Blvd, Old Westbury, New York 11568, USA

*Corresponding author: bbeatty@nyit.edu

**Abstract**:

Atherosclerotic lesions within carotid and cerebral vessels are likely to influence hemodynamics and manifest into vascular pathologies, including Alzheimer's Disease and ischemic stroke. Hemodynamics are influenced by changes in luminal diameter of vessels and wall shear stress derived from turbulence, which directly relates to the surface topography of the lumen. In this study, we performed a quantitative assessment of surface metrology of carotid and cerebral arteries in relation to vessel size and location among individuals. We speculate intracranial vessels will follow suit of extracranial vessels, with increased surface roughness in larger-diameter vessels.

Samples of the internal carotid, common carotid, and multiple branches of the Circle of Willis were collected at 18 different sites from 10 human whole body donors. The arterial surface metrology was analyzed using a Sensofar S Neox 3D optical profiler, from which ISO 25718-2 areal roughness parameters, texture direction, motifs analysis, and scale-sensitive fractal analyses were analyzed using SensoMap software.

The most significant differences between individuals, though surface roughness appears also greater in the larger vessels. In comparison, the side (left vs. right) is almost immaterial. With further research in this field, the pathophysiology of intracranial atherosclerosis a the role of atherosclerosis in neurodegenerative disorders.

**Keywords**: surface metrology, arteriosclerosis, atherosclerosis, cerebral artery, Circle of Willis, intracranial plaque

**Introduction**:

Atherosclerosis is a multifocal, arterial disease defined as lipid accumulation within the intimal vessel layer that progresses into thickening and stiffening of a vessel wall. The pathophysiology behind this disease involves repetitive endothelial cell dysfunction induced by hemodynamic demands within arterial vessels (Chiu and Chien, 2011). Blood flow and hemodynamic forces throughout the vascular system are not uniform and explain the disparities among patterns of endothelial damage in various locations. Numerous studies (Hoehmann et al. 2017; Chiu and Chien, 2011; Wong et al., 2020) demonstrated that atherosclerotic lesions preferentially develop at arterial branches with turbulent blood flow and low wall shear stress, in contrast to straight arterial vessels that exhibit laminar flow and high shear stress. Furthermore, such turbulent flow and low shear stress upregulate atherogenic genes in endothelial cells that further promote atherogenesis (Chiu and Chien, 2011). Thus, a vicious cycle emerges between surface roughness and turbulent flow, as atherogenic plaques promote a jagged luminal surface that encourages turbulent flow and predisposes the individual to vascular pathologies.

Throughout recent years, there has been extensive research dedicated to patterns of atherosclerosis. Certain cardiovascular risk factors promote progression of atherosclerosis, including age, genetics, tobacco smoking, hypertension, and diabetes (Chiu and Chien, 2011). In general, it has been established that atherosclerotic lesions are more often located in larger vessels at areas with turbulent flow and low shear stress. However, there is little consensus as to what constitutes a large vs. small vessel. Some studies categorize vessels based on diameter size, while others compare based on location (proximal vs. distal) (Aboyans et. al, 2007). Larger vessel disease may develop as early as late teenage years, while smaller vessel disease develops in later decades of life. The lack of advanced plaque formation in smaller intracranial vessels may be attributed to the decrease in blood pressure and vascular resistance as extracranial arteries feed into the Circle of Willis (Denswil et. al 2016). Despite this intrinsic protective mechanism of the intracranial vessels, as the larger vessels stiffen and lose their elasticity with aging, it is assumed to increase intracranial vascular resistance and initiate the cycle of surface damage and turbulent flow. In addition, large vs small vessel disease has been associated with atherosclerotic risk factors. Smoking and hypercholesterolemia are more significantly associated with large vessel disease, while diabetes is more significantly associated with small vessel disease (Aboyans et. al, 2007). Hypertension was also found to specifically contribute to intracranial small vessel lesions. These intracranial artery lesions are correlated with ischemic stroke, cognitive decline, and vascular events (Wu et al., 2018).

It is clear that atherosclerosis of an artery narrows the lumen and limits adequate blood flow within the vessel, a process that has been linked to cognitive decline and Alzheimer's disease (AD). A study comparing AD patients with age-matched healthy controls revealed that AD patients had more severity and number of atherosclerotic plaques in the cerebral arteries (De

Eulate et. al, 2017). The mechanism behind AD and plaques was explained by the occlusion-induced decrease in both oxygen and glucose, which then results in a change in the cerebral homeostasis leading to an increase in oxidative stress, inflammation, and dysfunction of neurotransmitters.

Calcifications of the cerebral arteries are also known to be correlated with ischemic stroke. In fact, carotid atherosclerosis is the underlying cause of the majority of strokes (Wu et al, 2018). Despite established evidence of thromboembolism as the major cause of ischemic stroke, there is insufficient research on the trajectory of emboli based on hemodynamic patterns, largely due to the complexity of the cerebrovascular system. Therefore, analysis of atherosclerosis patterns through surface metrology of the arterial lumen holds considerable importance in understanding vascular pathology.

Research has shown that arterial luminal surface roughness can be used to identify various diseases within different biological tissues. Measurement of the carotid intima-media thickness is widely used as a marker for identifying atherosclerosis. Niu et. al (2013) illustrated that the endothelium luminal surface increases in roughness before increasing in thickness, suggesting analysis of the luminal surface as a potential method of early diagnosis of atherosclerosis. Yet the measures of roughness by Niu et al (2013) were limited to a coarse measurement of roughness based on high-resolution ultrasound data, prior to the advent of widespread availability of non-contact areal roughness measures. More recent studies of intraplaque hemorrhage using MRI (Lu et al., 2019) capture shape data, but cannot measure how such plaques or changes in luminal topography might change hemodynamics with data that could be integrated into flow simulations. Wong et al., (2020) achieve simulations of blood flow through real vessels, though the scale of the source measurements are limited to vessel curvature, diameter, and branching, and cannot include the influence of microtopographical surface characteristics., such as roughness, that can have a great impact on wall shear stresses and flow turbulence.

Engineers studying the tribology of machines and tools had sought ways to measure surface roughness for decades (Jiang et al, 2003), and in 2012 the International Standardization Organization formalized a set of areal roughness parameters, ISO-25718-2 for non-contact surface metrology. Since then, an increasing number of applications of these measures of biological surfaces have enabled a number of quantitative assessments of surface roughness of teeth (Mihlbachler et al., 2022), bone (Hoehmann & Beatty, 2022), and skin (Ohtsuki, et al., 2013).

To further the understanding of atherosclerotic changes and cerebrovascular pathologies, the endothelial surface roughness of carotid and cerebral arteries were quantified. Numerous tissue samples were collected from the Circle of Willis and carotid arteries of both sides of ten cadavers to analyze the significant differences and similarities between the specific vessels, vessel size, vessel side (left vs. right), and between individuals. Ideally, asymmetry in luminal roughness should not be found, as there is no evidence of asymmetry in atherosclerosis generally. In previous studies, vessel size is positively associated with wall shear stress and peripheral vascular resistance, leading to increased predisposition to endothelial damage and vessel formation. We expect to have similar outcomes. More than anything else, it is predicted that the surface roughness would be most dependent upon individual differences, as the population of donors in gross anatomy labs tend to exhibit a wide range of atherosclerosis severity (Schear & Beatty, 2014; Hoehmann et al., 2017), and can be attributed to a variety of health, behavioral and genetic factors.

**Materials and Methods**

Sample Collection: Ten cadavers from the NYITCOM Anatomy Lab that were preserved with formaldehyde and phenol were used for this study. Donors were assigned a number to protect their identity. The age (65-94 years old), gender (five male, five female), cause of death (as reported on their death certificate), and vessels sampled for each donor used in this study can be found in Table 1.

Tissue Collection: Ten cadaver brains were dissected for sample collection. A five millimeter sample of each of the following Circle of Willis arteries were extracted using dissecting scissors: basilar, left/right vertebral, left/right superior cerebellar, left/right posterior cerebral, left/right posterior communicating, left/right middle cerebral, left/right anterior cerebral, and anterior communicating artery (Figure 1). The internal carotid artery and common carotid arteries were additionally dissected from the ten cadavers (Figure 1). The samples were stored in specimen containers. Some of the arteries were inadequate and/or tainted and therefore not used during analysis, therefore the number of the artery samples collected from each individual varied.

Sample Preparation: A 5 mm arterial tissue sample was cut longitudinally using dissecting scissors and mounted on glass slides with the inner luminal surface face up for scanning. The samples were secured onto the slide by taping down the edges and ensuring the center of the sample surface was flattened for proper scanning results. Any visible blood or contaminants were gently removed using forceps or water as needed. An effort was made to have minimal touch contact with the luminal surface in order to minimize surface contamination or damage. A minimum of two sample slides were prepared for each artery.

Scanning Procedure: The luminal surface of all the arterial samples were scanned using a Sensofar S Neox 3D optical profiler in confocal fusion mode. All of the samples were scanned using the same procedure and utilizing the SensoSCAN software program. The slides were handled and placed on the stage with care to avoid any artifacts. First, a lower magnification of either x 5 or x 20 was used to orient the observer and find the appropriate scanning location. The observer would locate a relatively flat area of the arterial sample for optimal scanning. After the region was chosen, the objective was set to 20x, the z range was defined by what was in focus, and the scan was performed. All scans were an area of 873.33 x 656.61 micometers, comprising 1354 x 1018 pixels (pixel size = 0.64 micrometers). Scans missing 10% or more of data were rejected and the sample was rescanned until an adequate scan was obtained.

Post Processing Scan Data: Surface metrology files were then imported to the SensoMAP software program for data analysis and visualization. All of the scans were leveled, form removed (polynomial 8), retouched, and filling in of non-measured points (when needed). Removing form is intended to reduce unintended effects of measuring warped surfaces. Retouching deletes contaminants from the scan, such as dust particles and sources of unusual reflections. Filling in non-measured points replaces gaps with a smooth surface confluent to surrounding points to create a solid surface. Gaps are areas not scanned (usually because confocal scanning has limited ability to collect data from steep angled surfaces to the horizontal plane) or regions deleted during retouching. Lastly, a second leveling operator was applied. After processing, several studies were performed, including a 3D view, motifs analysis, scale-sensitive fractal analysis (SSFA), and texture direction. These data types are depicted in Figure 2, and a full list of parameters studied here can be found in Table 2.

Statistical Analysis:

To determine what sorts of tests are appropriate, a basic test of normality was done in SPSS (Version 28, IBM). If normal, one-way ANOVAs will be run for those hypotheses that have more than 2 groups. These included questions of the influence of side (left, right, and midline), vessel (9 vessels named above), vessel size (large, medium, small, and tiny), and individual (10 individuals)  To examine what relationships were the sources of significant variances in means, Tukey's post-hoc test was done.

Furthermore, though data were collected for ISO 25718-2, texture direction, motifs analysis, and SSFA, multiscale analyses such as SSFA are arguably best for characterization of surfaces. ISO 25781-2, texture direction, and motifs analyses are best used for detailed descriptions of surface features at expected scales, making them potentially most useful once a more significant body of surface metrology data is accumulated for a study system. Therefore, here we will focus our discussion on the SSFA parameters to establish differences between study variables, though we report all data collected (see the Supplemental Materials for the complete dataset).

**Results**

Test of normality

The majority of variables measured in this study exhibited normal distributions. This is not surprising, as most analyses of such data turn out this way, for a variety of studies from teeth (Mihlbachler et al., 2022) to bones (Hoehmann & Beatty, 2022). In such studies, parametric and nonparametric analyses typically yield identical or very similar results, so these data were analyzed with a one-way ANOVA to test for variance in their means like those of bones and teeth mentioned above.

ANOVA Side

The ANOVA for side tested differences in variance between left and right sides, as well as with the single vessel that belongs to neither side, the basilar artery. Among SSFA parameters, only smooth-rough crossover (SRC) significantly differed in variance between the sides, specifically left and right sides (Figure 3). In addition, multiple other parameters measured were significantly different between these groups (Supplemental material), and the post-hoc Tukey test allowed identification of where such differences could be found. More parameters differed between left and right sides (Vm, Vmp, Sda, Sdv, Spk) than left versus basilar (Ssk, Sku, Smr, & # Motifs) or right versus basilar (Sku & Smr). No single parameter differed significantly between all three groups. Overall, this suggests that comparatively few of the parameters differed significantly between left and right sides. This is somewhat reassuring, possibly indicating that surface metrology was not affected by random influences of asymmetry.

ANOVA Vessel

The ANOVA for vessel combined the data for each vessel regardless of side, but maintaining their individual branch identity. This allows the identification of differences between branches when subject to a post-hoc Tukey test.

SSFA parameters differ significantly in their means between only four of the 9 vessels sampled (Figure 4). These include between the anterior cerebral artery and three others, internal carotid (Y Max an SRC threshold), common carotid (Y Max, SRC threshold), and vertebral artery (Y Max, SRC threshold, Lsfc, and DIs). In addition, the superior cerebellar artery differs from the same three vessels significantly: internal carotid (Y Max an SRC threshold), common carotid (Y Max, SRC threshold, Lsfc), and vertebral artery (Y Max, SRC threshold, Lsfc, and DIs). Though these two vessels do not fit within a specific size category together, it is notable that they both differ significantly from the three largest arteries included in this study.

Multiple other parameters were found to be significantly different between vessels (Supplemental materials).

ANOVA Size

Vessels were divided into groups based on approximate sizes. Large includes internal carotid and common carotid. Medium includes basilar and vertebral arteries. Small includes all of the posterior, middle, and anterior cerebral arteries. Tiny includes the posterior communicating and superior cerebellar arteries. Among SSFA parameters, multiple parameters differed between vessel size categories (Figure 5). Y Max significantly differed between large and small, between large and tiny, and between medium and tiny vessels. Smooth rough crossover threshold significantly differed between large and small, large and tiny, and medium and tiny vessels. Smooth rough crossover (SRC) differed significantly between large and small vessels. Fractal complexity (Lsfc) differed between large and tiny, and medium and tiny vessels. Fractal dimension (DIs) differed between large and tiny, and medium and tiny vessels. Multiple other parameters studied here have significant differences in their variance (Supplemental materials). Tukey's post-hoc test indicates that these differences are found significantly between large and small vessels (26 parameters), large and tiny vessels (16 parameters), and medium and tiny vessels (7 parameters).

ANOVA Individual

Between individuals, only two SSFA parameters differed significantly, specifically $R^2$ coefficient and scale of maximum complexity (Smfc, Figure 6). $R^2$ coefficient differed between individuals 4 and 7, 1 and 4, 7 and 10, 1 and 8, and 1 and 10. Smfc only differed between individuals 2 and 8. It is worth noting here that cause of death information for two of these individuals, 8 and 9, included diagnosis of Parkinson's disease. These sample sizes of neurodegenerative disorders are not great, and further sampling, perhaps with more granular details of medical history, are necessary to derive any analytical value from this anecdotal observation.

Multiple other parameters measured in this study, including ISO 25718-2 and motifs analysis parameters, demonstrated significantly different variance in means between individuals (Supplemental materials). Thus, when looked at together, these data indicate that individual differences outweigh the impact of side, size, and specific vessel branch.

**Discussion**

Our research determined the significant contributing factors between variance in surface roughness, isotropy, motifs analysis, and scale-sensitive fractal analysis are attributed to differences between individuals themselves and the vessel size. In comparison, the side (left vs. right) is almost immaterial. Perhaps the most compelling aspect of these results to note is the restricted range of variance among the larger vessels (common carotid, basilar, internal carotid, and vertebrals) and the larger range of variation in the data from smaller vessels. The coherence of large-diameter vessel topography indicates similar endothelial adaptations and degree of plaque amongst individuals, in comparison to the smaller-diameter vessel topography discoherence. This may suggest that intracranial small vessel disease is more influenced by individual risk factors, while the larger vessel disease is influenced by risk factors and more influenced by hemodynamic factors.

Individuals with more atherosclerotic risk factors are more inclined to plaque development. Our results support a prior study analyzing atherosclerosis within the Circle of Willis, which concluded that plaque formation was dependent upon differences between individuals themselves, such as the individual's age, smoking status, and presence/absence of diabetes mellitus (Denswil et al., 2016). Additionally, it showed plaque formation may occur anywhere among the Circle of Willis, but early and advanced lesions are more abundant in larger diameter vessels, which is supported by our data as well. A meta-analysis on the previous autopsy studies performed by Aboyans et. al, (2007) revealed significant contribution of age, diabetes, hypertension, and smoking on the progression of atherosclerotic disease. More importantly, it demonstrates that the cardiovascular risk factors are "not uniformly associated to atherosclerosis developed at either site" (Aboyans et. al, 2007). In fact,the study demonstrated a linear relationship between age and larger extracranial vessel disease, compared to the exponential relationship between age and smaller intracranial vessel disease and age. Additionally, smoking and hypercholesterolemia was found to contribute to larger vessel disease, while diabetes contributes more smaller vessel disease and hypertension is more with distal intracranial lesions. Despite the lack of detailed background of the cadavers, our study bolsters these findings as we displayed a variation of surface topography amongst individuals, portraying the non-uniform distribution of atherosclerotic lesions resulting from individual risk factors. The comorbid health diseases, genetics, and smoking status in combination with the histology of the arterial samples, would have created a more complete representation of the individuals' health and could be used to identify the personalized risk factors impacted atherosclerotic disease .

These results are encouraging, as it suggests that there are preferential patterns of atherosclerosis relating to size and specific arterial branch observed, and moreover that individuals with differences in health status may be recognized with surface metrology. The key remaining question of our study is the health status of these individuals, which may be best answered by histopathological approaches. Furthermore, in addition to classic histopathological approaches with H&E stained tissues that can illustrate AHA arteriosclerosis classes (Stary, 2000; Hoehmann et al., 2017), alizarin red staining would allow for study of calcification, which is presumably the underlying reason for changes in topography (Koscziucek et al., in review). It is unclear whether early atherosclerotic stages preceding calcification would result in topographical changes in the lumen, but the study of H&E stained samples of these vessels in comparison with these data may illuminate some subtle changes that might be overlooked by less sensitive methods such as microCT.

Conventional imaging techniques used for atherosclerosis identification include ultrasound, computed tomography, and magnetic resonance imaging. Despite the technological advancements in these methods to identify objective and qualitative atherosclerotic plaques, there are limitations. For example, nonenhanced CT is only able to detect calcifications, which is just one contributing factor to atherosclerotic plaques (Bos, et al, 2015). Bright (B) mode ultrasound doppler mode is used first-line to assess the carotid artery and other peripheral arteries as this technique offers a cost-effective, broad imaging that occurs in real-time, with the absence of radiation. Although it can be used to locate atherosclerosis with blood flow velocity, and a degree of plaque surface roughness, it has limitations in depth of examination and nascent plaques (Cismaru et al., 2021).

Despite the limited research in surface metrology of the cerebral arteries, there is extensive research of extracranial vessels to create inclusion criteria that diagnose atherosclerosis stages. One study defined a "vulnerable" plaque as a lesion with a fibrous cap <65 μm thick based on the actual thickness of the histologic sections from measurements made of plaque ruptures (Virmani et al., 2003). The earliest histological sign of atherosclerosis is the lipid-filled macrophages within the intima layer of the artery, creating the "fatty streak" (Niu et al., 2013). This research of the surface analysis within coronary arteries have eluded that surface changes occur even before intimal thickening, suggesting that more detailed surface topography studies can be used for diagnosing silent disease progression that would be missed on ultrasound.

Surface topography of arteries combined with patient demographics, comorbidities, genetics, and health behaviors, can be used to classify surface characteristics of developing atherosclerosis and predict symptomatology. Ideally, such additional information post-mortem along with surface metrology would allow us to investigate the progression of atherosclerosis in a more nuanced way that, if correlated with techniques typically used in diagnostics of living patients, could lead to a more accurate prognosis. Early detection and intervention of atherosclerosis can delay or potentially prevent the development of stroke, dementia, heart attacks and other vascular pathologies, and if these data can eventually be correlated to ultrasound and other methods used for living patients, improving early detection may be possible.

**Limitations of the study**

The limitations of the present study should be considered. The biggest challenge was the varying number and type of arteries sampled from each individual, as some arteries were damaged upon dissection. This may have underestimated the differences among each individual artery, relative side of Circle of Willis, or individuals. Moreover, the utility of these data would be amplified if coupled with techniques more commonly used in diagnoses of atherosclerosis in living patients. Though ultrasound is not utilized to investigate atherosclerosis in cerebral vessels, one could perform ultrasound assessments on the carotids of the same cadavers to compare these data types. Such a study would not work on cadavers preserved with formalin and phenol, as medical school labs typically use (as in this study), but ultrasound may work on reperfused Thiel-embalmed cadavers (Niels, 2022). Due to unavailability of reperfused Thiel-embalmed cadavers, such data collection was not in the scope of this study.

**Conclusion**

Overall, this current study demonstrated atherosclerotic surface changes of the arterial lumen of the cerebral vessels and the carotid vessels are dependent on the individual's risk factors and vessel size. As mentioned previously, recent research in this field determined age, genetics, smoking status, diabetes, and hypertension constitute the largest contribution in luminal surface roughness. Moreover, the study revealed differences between cerebral vessel size; larger diameter vessels exhibited more arterial luminal endothelial changes in comparison to smaller diameter vessels. Additionally, there was more topographic variation in smaller-diameter vessels between individuals, suggesting that small and large vessel diseases are not uniformly associated with the same risk factors. The vessel side held almost no significance. To further investigate atherosclerosis patterns and preferences, surface metrology research should be conducted with the corresponding histology of the samples to precisely visualize and confirm the surface changes, alongside ultrasound investigations of the same individuals (presumably Thiel-embalmed cadavers). All considered, such investigation should lead to more research of arterial surface adaptations as a result of individual cardiovascular risk factors in order to correlate luminal surface topography with health status in efforts to predict stroke, Alzheimer's Disease, and cognitive decline risk levels.

**Acknowledgements**

Kelsi Hurdle (NYITCOM Visualization Center) provided training and support for the Sensofar S Neox.

**Table captions**

**Table 1**. Samples from donors. The age, gender, cause of death, and the vessels collected is displayed in the table with relation to each cadaver. The cadaver's were designated number/letter identifiers to protect their identity.

|  | Cadaver Identity | | | | | | | | | | |
|---|---|---|---|---|---|---|---|---|---|---|---|
| Individual number | 1 | 2 | 3 | 4 | 5 | 6 | 7 | 8 | 9 | 10 | |
| In-house record number | 2020-147 | B185 | B154 | B228 | B250 | 2020-084 | 2020-167 | B231 | 2020-006 | 2020-051 | TOTAL |
| Age | 83 | 79 | 78 | 68 | 94 | 65 | 87 | 87 | 86 | 92 | |
| Gender | F | F | M | M | M | F | F | M | F | M | |
| Cause of Death | bronchoalveolar lung cancer | anal cancer | cardiopulmonary arrest, metastatic cancer, melanoma | cardiopulmonary arrest, congestive heart failure, cardiomyopathy | failure to thrive, severe protein calorie nutrition, dementia, sick sinus syndrome | fatal acute ventricular arrhythmia, probable MI | unknown to donor program | failure to thrive, Parkinson's disease, chronic kidney disease, chronic obstructive pulmonary disease | Parkinson's disease | cardiac arrhythmia, failure to thrive | |
| **Basilar Artery** | ✓ | ✓ | ✓ | ✓ | ✓ | ✓ | ✓ | ✓ | ✓ | ✓ | 10 |
| **Right Vertebral Artery** | ✓ | ✓ | ☐ | ✓ | ☐ | ☐ | ✓ | ☐ | ☐ | ✓ | 5 |
| **Left Vertebral Artery** | ✓ | ✓ | ☐ | ✓ | ☐ | ☐ | ☐ | ☐ | ✓ | ✓ | 5 |
| **Right Superior Cerebellar Artery** | ✓ | ✓ | ✓ | ✓ | ✓ | ✓ | ✓ | ✓ | ✓ | ✓ | 10 |
| **Left Superior Cerebellar Artery** | ✓ | ✓ | ✓ | ✓ | ✓ | ☐ | ✓ | ✓ | ☐ | ✓ | 8 |
| **Right Posterior Cerebral Artery** | ✓ | ✓ | ✓ | ✓ | ✓ | ✓ | ✓ | ✓ | ☐ | ☐ | 8 |
| **Left Posterior Cerebral Artery** | ✓ | ✓ | ✓ | ✓ | ✓ | ✓ | ✓ | ✓ | ☐ | ☐ | 8 |
| **Right Posterior Communicating Artery** | ✓ | ☐ | ✓ | ✓ | ✓ | ☐ | ☐ | ☐ | ☐ | ☐ | 4 |
| **Left Posterior Communicating Artery** | ✓ | ✓ | ☐ | ☐ | ☐ | ☐ | ☐ | ☐ | ☐ | ☐ | 2 |
| **Right Middle Cerebral Artery** | ✓ | ✓ | ✓ | ✓ | ✓ | ✓ | ✓ | ✓ | ✓ | ☐ | 9 |
| **Left Middle Cerebral Artery** | ✓ | ✓ | ✓ | ✓ | ✓ | ✓ | ☐ | ✓ | ✓ | ☐ | 8 |
| **Right Anteiror Cerebral Artery** | ✓ | ✓ | ✓ | ✓ | ✓ | ✓ | ✓ | ✓ | ✓ | ☐ | 9 |
| **Left Anterior Cerebral Artery** | ✓ | ✓ | ✓ | ✓ | ✓ | ✓ | ✓ | ✓ | ✓ | ☐ | 9 |
| **Anterior Communicating Artery** | ☐ | ☐ | ☐ | ☐ | ☐ | ☐ | ☐ | ☐ | ☐ | ☐ | 0 |
| **Right Common Carotid Artery** | ✓ | ✓ | ☐ | ✓ | ☐ | ✓ | ✓ | ✓ | ✓ | ☐ | 7 |
| **Left Common Carotid Artery** | ✓ | ✓ | ☐ | ✓ | ☐ | ✓ | ✓ | ✓ | ✓ | ☐ | 7 |
| **Right Internal Carotid Artery** | ✓ | ✓ | ☐ | ✓ | ☐ | ✓ | ✓ | ✓ | ✓ | ☐ | 7 |
| **Left Internal Carotid Artery** | ✓ | ✓ | ☐ | ✓ | ☐ | ✓ | ✓ | ✓ | ✓ | ☐ | 7 |
| TOTAL | 17 | 16 | 10 | 16 | 10 | 12 | 13 | 13 | 11 | 5 | |

**Table 2**. Parameters used in this study, including International Organization for Standardization (ISO) 25718-2 areal roughness parameters, texture direction, motifs analysis, and scale sensitive fractal analysis (SSFA).

| Name | Unit | Context | Description |
|---|---|---|---|
| ISO 25718-2 | | | |
| Height parameters | | | |
| Sq | µm | | Root-mean-square height |
| Ssk | <no unit> | | Skewness |
| Sku | <no unit> | | Kurtosis |
| Sp | µm | | Maximum peak height |
| Sv | µm | | Maximum pit height |
| Sz | µm | | Maximum height |
| Sa | µm | | Arithmetic mean height |
| Hybrid parameters | | | |
| Smr | % | c = 1 µm under the highest peak | Areal material ratio |
| Smc | µm | p = 10% | Inverse areal material ratio |
| Sxp | µm | p = 50% q = 97.5% | Extreme peak height |
| Functional parameters (Volume) | | | |
| Sal | µm | s = 0.2 | Autocorrelation length |
| Str | <no unit> | s = 0.2 | Texture-aspect ratio |
| Std | ° | Reference angle = 0° | Texture direction |
| Feature parameters | | | |
| Sdq | <no unit> | | Root-mean-square gradient |
| Sdr | % | | Developed interfacial area ratio |
| Functional parameters (Volume) | | | |
| Vm | µm³/µm² | p = 10% | Material volume |
| Vv | µm³/µm² | p = 10% | Void volume |
| Vmp | µm³/µm² | p = 10% | Peak material volume |
| Vmc | µm³/µm² | p = 10% q = 80% | Core material volume |
| Vvc | µm³/µm² | p = 10% q = 80% | Core void volume |
| Vvv | µm³/µm² | p = 80% | Pit void volume |
| Feature parameters | | | |
| Spd | 1/µm² | pruning = 5% | Density of peaks |
| Spc | 1/µm | pruning = 5% | Arithmetic mean peak curvature |
| S10z | µm | pruning = 5% | Ten point height |
| S5p | µm | pruning = 5% | Five point peak height |
| S5v | µm | pruning = 5% | Five point pit height |
| Sda | µm² | pruning = 5% | Mean dale area |
| Sha | µm² | pruning = 5% | Mean hill area |
| Sdv | µm³ | pruning = 5% | Mean dale volume |
| Shv | µm³ | pruning = 5% | Mean hill volume |
| Functional parameters (Stratified surfaces) | | | |
| Sk | µm | Gaussian filter 0.08 mm | Core roughness depth |
| Spk | µm | Gaussian filter 0.08 mm | Reduced summit height |
| Svk | µm | Gaussian filter 0.08 mm | Reduced valley depth |
| Smr1 | % | Gaussian filter 0.08 mm | Upper bearing area |
| Smr2 | % | Gaussian filter 0.08 mm | Lower bearing area |
| Spq | <no unit> | Gaussian filter 0.08 mm | Plateau root-mean-square roughness |
| Svq | <no unit> | Gaussian filter 0.08 mm | Valley root-mean-square roughness |
| Smq | <no unit> | Gaussian filter 0.08 mm | Material ratio at plateau-to-valley transition |
| Texture direction | | | |
| Isotropy | % | | Uniformity of orientation of tiles |
| First Direction | ° | | Most common orientation of tiles |
| Second Direction | ° | | Second-most common orientation of tiles |
| Third Direction | ° | | Third-most common orientation of tiles |
| Motifs analysis | | | |
| Number of motifs | <no unit> | | Number of recognized local peaks delimited by slope inflections |
| Height[Mean] | µm | | Mean height of motifs |
| Area[Mean] | µm² | | Mean area of motifs |
| Scale-Sensitive Fractal Analysis (SSFA) | | | |
| Y Max | <no unit> | | The size of tiles at the finest scale |
| SRC threshold | <no unit> | | Threshold relative length of the smooth-rough crossover |
| Smooth-rough crossover (SRC) | µm | | Scale of the smooth-rough crossover |
| Reg. coefficient R² | <no unit> | | A measure of the accuracy of the complexity value |
| Fractal complexity (Lsfc) | <no unit> | | How much of the surface is more complex than a Euclidean plane |
| Fractal dimension (Dls) | <no unit> | | The range of scales in which the slope of the line is straight |
| Scale of max complexity (Smfc) | µm | | Scale where highest complexity is found |

**Figures**

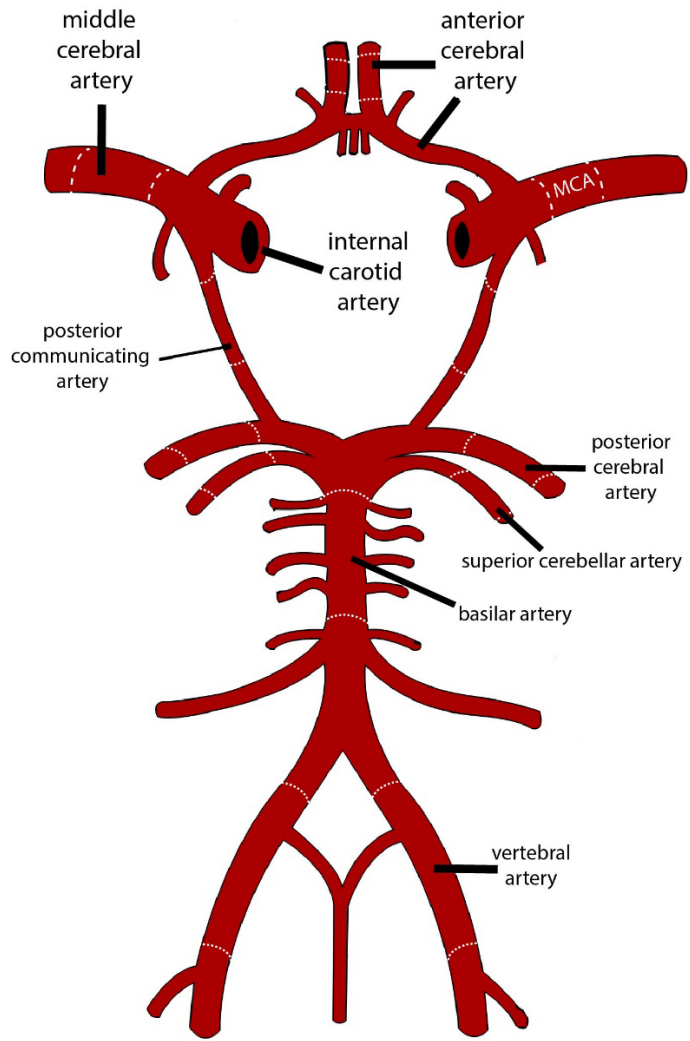

**Figure 1**. Arterial sampling. A 2D illustration of the Circle of Willis to define the precise location of where each vessel sample was dissected. Not pictured is the dissection location of the common carotid artery, which was sampled just distal to the bifurcation into external and internal carotids.

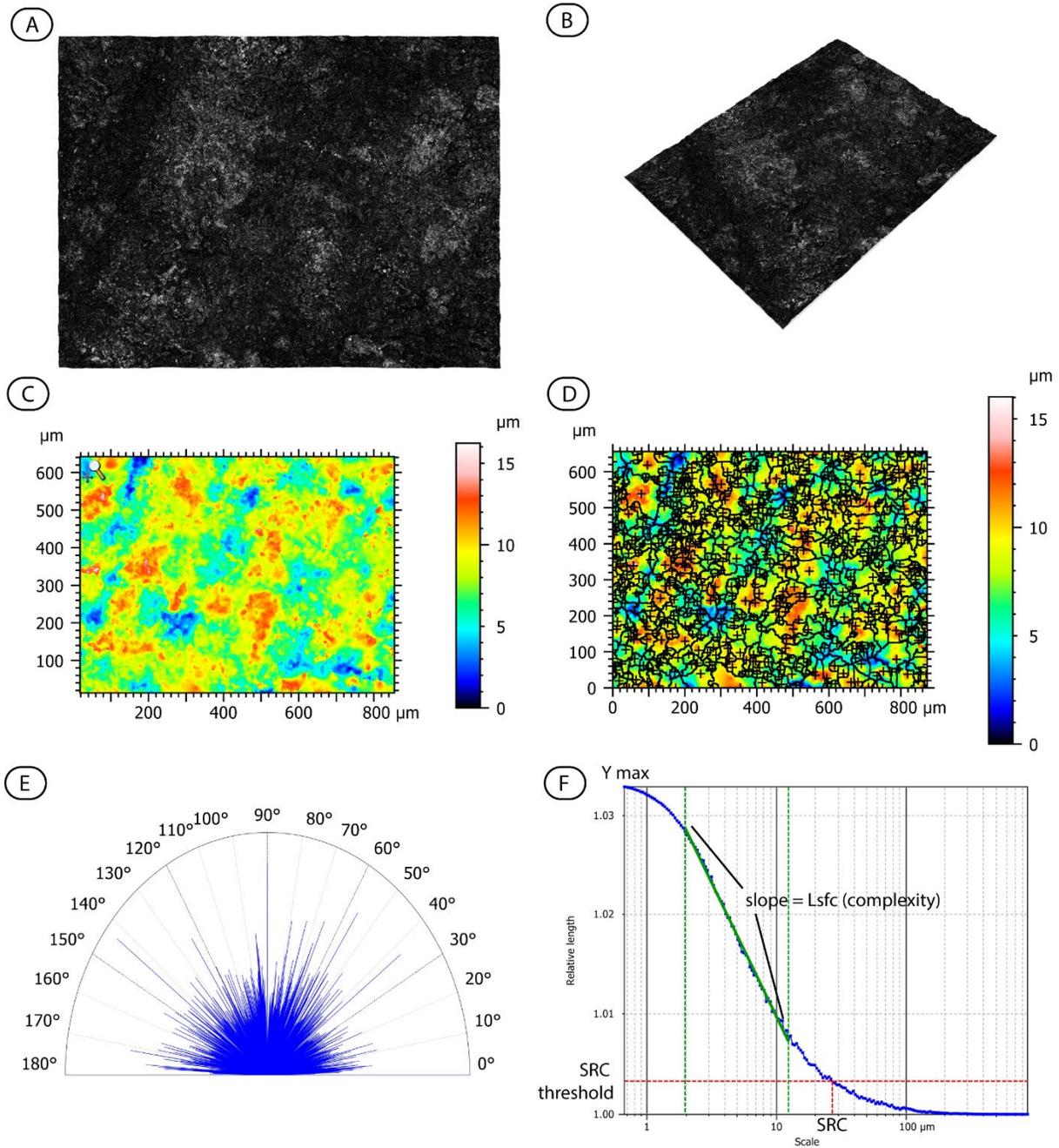

**Figure 2**. Examples of A) a scanned vessel (3D view); B) 45 degree angled view of 3D view to better view topography; C) heatmap of elevations of the scanned surface; D) motifs analysis; E) texture direction rose diagram; and F) scale-sensitive fractal analysis. Not pictured here are the ISO 25718-2 parameters derived from the surface scan (see Table 2).

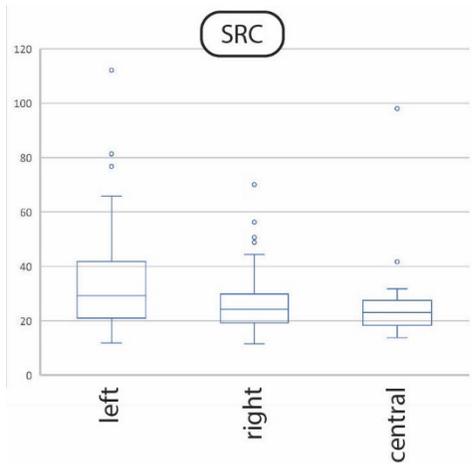

**Figure 3**. Scale-sensitive fractal analysis parameters exhibiting significant differences in means between sides (left, right, and central), as determined by ANOVA.

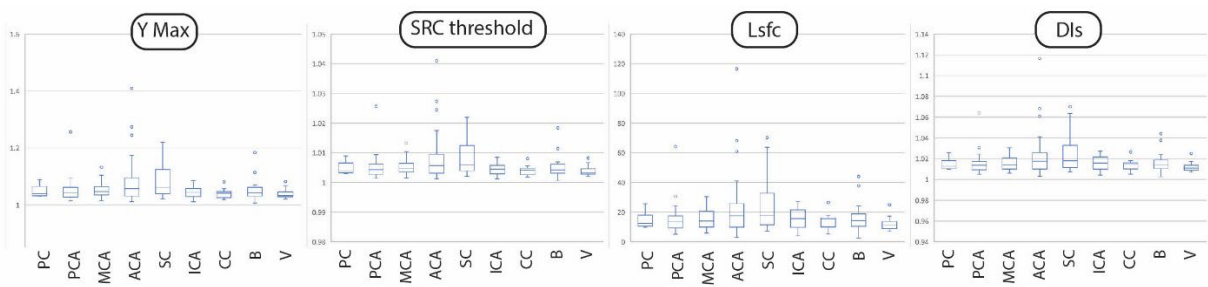

**Figure 4**. Scale-sensitive fractal analysis parameters exhibiting significant differences in means between vessels, as determined by ANOVA.

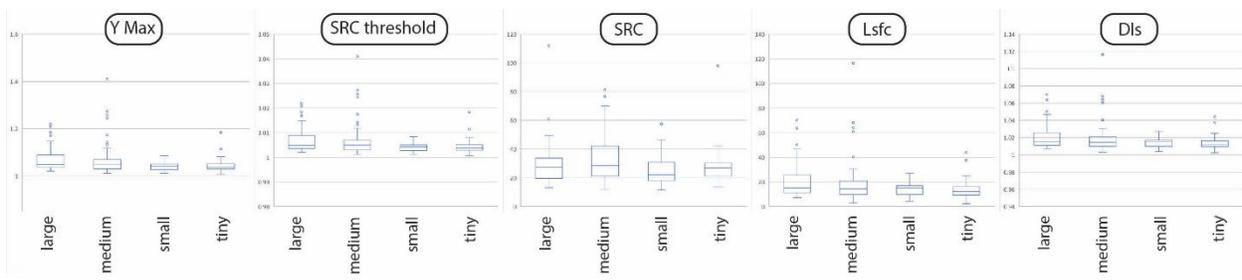

**Figure 5**. Scale-sensitive fractal analysis parameters exhibiting significant differences in means between size categories of vessels (large, medium, small, and tiny), as determined by ANOVA.

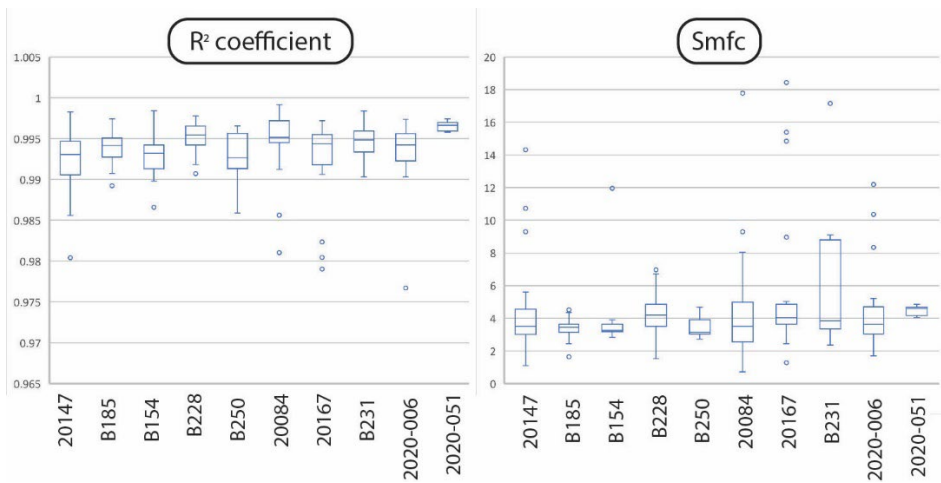

**Figure 6**. Scale-sensitive fractal analysis parameters exhibiting significant differences in means between individuals, as determined by ANOVA.

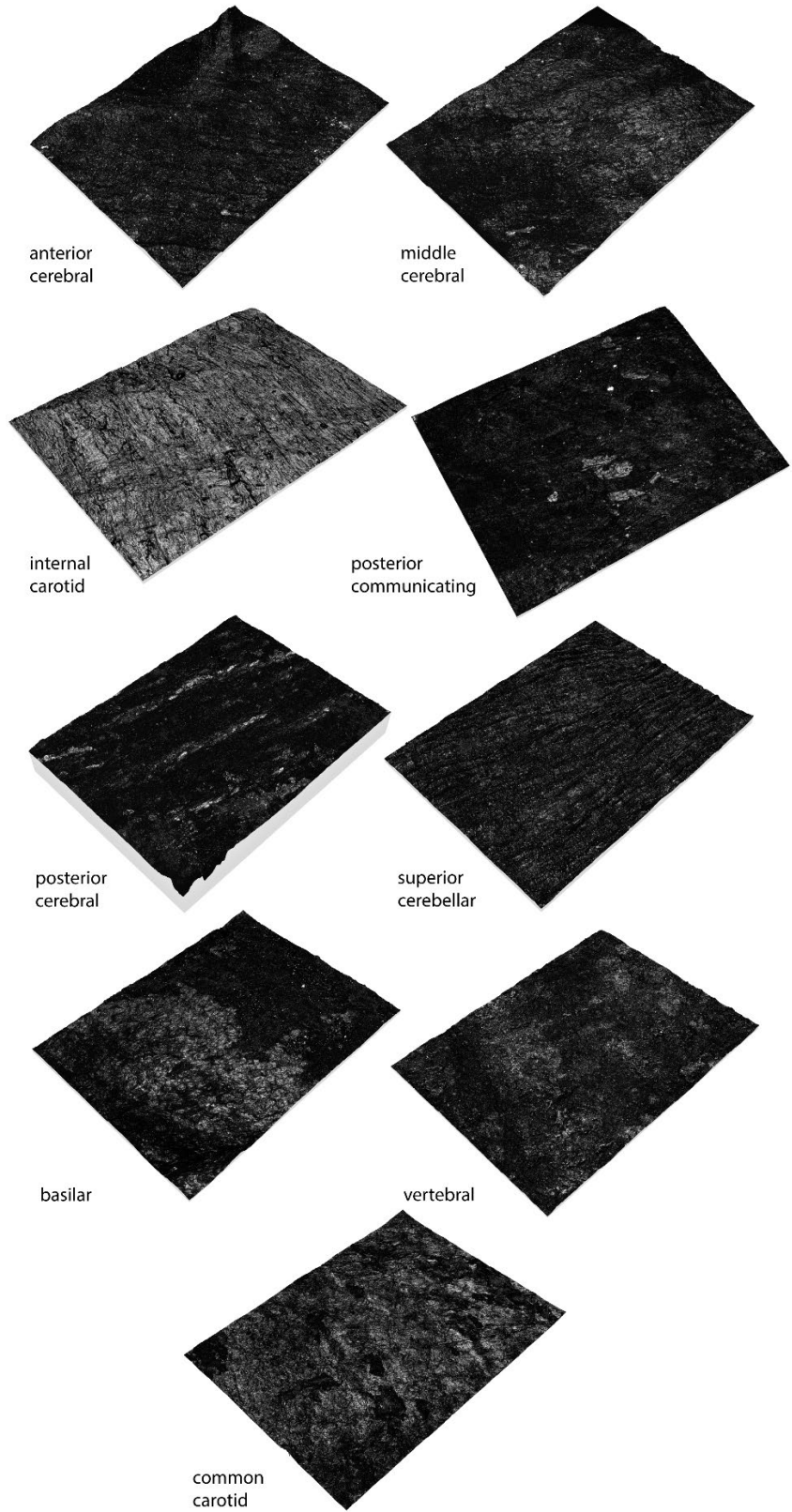

**Figure 7**. Oblique 3D views of the surface metrology scans of vessels.

**Supplemental Materials**

S1. The complete dataset of parameters derived from scans.